# Dynamical Spin Response Functions for Heisenberg Ladders

Danny B. Yang[1,2] and W. C. Haxton[2]

[1]*Department of Physics, Harvard University, Cambridge, MA 02138*
[2]*Institute for Nuclear Theory, Box 351550, and Department of Physics, Box 351560, University of Washington, Seattle, Washington 98195*

(September 1, 2018)

## Abstract

We present the results of a numerical study of the $2 \times L$ spin-$\frac{1}{2}$ Heisenberg ladder. Ground state energies and the singlet-triplet energy gaps for $4 \leq L \leq 14$ and $J_\perp / J_\parallel = 1$ were obtained in a Lanczos calculation and checked against earlier calculations by Barnes *et al.* (even $L \leq 12$). A related moments technique is then employed to evaluate spin response functions for $L = 12$ and a range of $J_\perp / J_\parallel$ (0 - 5). We comment on two issues, the need for reorthogonalization and the rate of convergence, that affect the numerical utility of the moments treatment of response functions.

Typeset using REVTEX

Heisenberg spin ladders have attracted considerable attention recently [1] due to possible connections to materials exhibiting high $T_c$ superconductivity: theoretical studies [2] hint at the possibility of even-chain ladders becoming superconductors when doped with charge carriers. There is some experimental support for this possibility as $Sr_{0.4}Ca_{13.6}Cu_{24}O_{41.84}$, a material with spin-$\frac{1}{2}$ chains and 2-chain ladders, was shown to superconduct at 12 K and 3 GPa [3].

Spin ladders are also fascinating theoretically because of their unexpected behavior when viewed as interpolators between the spin-$\frac{1}{2}$ 1D antiferromagnetic Heisenberg chain and the 2D square analog. The latter is fully ordered at low temperatures [4], while Bethe [5] demonstrated in the 1930's that spin-spin correlations in the 1D chain have a slow power-law decay. Yet the transition between these limits by forming spin-$\frac{1}{2}$ ladders with increasing numbers of legs is not smooth: ladders with even numbers of legs have a finite gap to the lowest triplet state and an exponential decay of spin-spin correlations, while odd-leg ladders have gapless excitations and a power-law fall off of spin-spin correlations [6]. These issues and examples of materials exhibiting these properties are discussed in several recent reviews (see, e.g., [1]).

In the hope of gaining deeper insight into such systems, numerical modelers have employed a variety of techniques to study spin ladders, including exact diagonalizations with the Lanczos algorithm [7], quantum Monte Carlo simulations [7,3], and approximate density matrix renormalization group methods using rung or plaquette bases (see White *et al.* [2] and Piekarewicz and Shepard [8]). The exact calculations, while limited to small $L$, play an important role in testing approximation schemes, and also in evaluating dynamic quantities, such as spin responses, that are difficult to treat in other approaches. In this report we present Lanczos results for the ground-state energy and singlet-triplet gap for $2 \times L$ systems through $L = 14$, which we compare to the even $L \leq 12$ calculations [7] of Barnes *et al.* We then study the evolution of the dynamical spin response function for $L = 12$ as the ratio of rung to leg interaction strengths $J_\perp/J_\parallel$ is varied. As this response can be measured in inelastic neutron scattering experiments [9], it provides an important test of spin interactions in ladder materials. This response has also been evaluated in recent plaquette basis approximation schemes [8]. The present calculations are based on a Lanczos moments expansion that can be iterated to arbitrary accuracy. We discuss some numerical aspects of this procedure that are relevant to spin ladder calculations.

The Hamiltonian for the spin-$\frac{1}{2}$ Heisenberg spin ladder consisting of two coupled chains is

$$H = J_\perp \sum_{\langle i,j \rangle_\perp} \vec{S}_i \cdot \vec{S}_j + J_\parallel \sum_{\langle i,j \rangle_\parallel} \vec{S}_i \cdot \vec{S}_j, \tag{1}$$

where $i$ is a lattice site on which one electron sits, $\langle i,j \rangle_\perp$ denotes nearest neighbor sites on the same rung, and $\langle i,j \rangle_\parallel$ denotes nearest neighbors on either leg of the ladder. We used periodic boundary conditions along the legs of the ladder. The ratio $J_\perp/J_\parallel$, the relative strength of the rung and leg interactions, depends on the choice of material being modeled. In the strong coupling limit, $J_\perp/J_\parallel \to \infty$, the electrons across each rung form an $S = 0$ pair, and the ground state wave function is the resulting product [7]. The ground state thus has $S = 0$ and an energy/spin proportional to $J_\perp$, with perturbative corrections of relative size $J_\parallel/J_\perp$.

In the Lanczos algorithm [10] the Hamiltonian is written in tridiagonal form recursively, using a series of operations

$$H|v_i\rangle = \beta_{i-1}|v_{i-1}\rangle + \alpha_i|v_i\rangle + \beta_i|v_{i+1}\rangle \qquad (2)$$

in which the next basis vector $|v_{i+1}\rangle$ is generated from the previous one, $|v_i\rangle$, with the choice of $|v_1\rangle$ depending on the application. We truncated this series after $k$ steps and diagonalized the resulting $k \times k$ matrix by the QL algorithm [12]. The resulting energy per spin for the singlet ground state and gap to the first triplet state are compared to the results of Barnes *et al.* in Tables I and II. The agreement is very good, with only very minor differences appearing for large $L$. Our calculations through $L = 13$ were done in both single and double precision: the results are identical to the accuracy employed in the tables. The $L = 14$ calculations were performed in single precision only.

The ground-state energy per spin can be extrapolated to the bulk limit using a scaling function similar to that of Barnes *et al.* [7]

$$f(L) - f(\infty) = C_0(-1)^L \frac{e^{-L/L_0}}{L^p} \qquad (3)$$

where $p = 2$. A fit to the results of Table I yields $C_0 = 1.12$, $L_0 = 3.86$, and $f(\infty) = $ -0.5780. The even-$L$ results for the spin-triplet gap extrapolate with $p = 1$ to $f(\infty) = 0.502$ ($C_0 = 3.61$, $L_0 = 3.82$).

We now consider the evolution of the dynamic spin response function $S(\vec{q}, \omega)$, where $\vec{q}$ and $\omega$ are the three-momentum and energy transfer, as $J_\perp/J_\parallel$ is varied. $S(\vec{q}, \omega)$ is defined by

$$S(\vec{q}, \omega) = \sum_n |\langle n|\vec{S}(\vec{q})|g.s.\rangle|^2 \delta(\omega - \omega_n) \qquad (4)$$

where $|g.s.\rangle$ denotes the ground state and where the sum is taken over a complete set of excited states $|n\rangle$ of energy $\omega_n$. The spin transition operator is

$$\vec{S}(\vec{q}) = \sum_j \vec{S}_j e^{i\vec{q}\cdot\vec{r}_j}, \qquad (5)$$

where the sum extends over all sites. In particular, if the dynamic spin response is probed at $q = (q_x, q_y) = (\pi, \pi) = \vec{q}_{\pi\pi}$, in units of the inverse lattice spacing, then

$$\vec{S}(\vec{q}_{\pi\pi}) = \sum_j (-1)^j \vec{S}_j, \qquad (6)$$

so that the operator sign alternates from site to site.

In the Barnes *et al.* work Lanczos techniques were employed in evaluating $S(\vec{q}_{\pi\pi}, \omega)$ for L=8 [7]. Quite recently Piekarewicz and Shepard [8] studied the $L = 6, 8$, and 16 systems by exploiting a plaquette truncation of the basis. Thus one motivation for the present effort is to provide a series of exact calculations in somewhat larger systems ($L$=12) that can serve as benchmarks for approximate methods, like those of Piekarewicz and Shepard, that are now being applied to dynamical quantities.

The Lanczos method is particularly well suited to the evaluation of inclusive response functions. Once an initial Lanczos expansion has been carried out to the point where the ground state is fully converged, the vector

$$S_z(\vec{q}_{\pi\pi})|g.s.\rangle \tag{7}$$

can be formed and its norm determined. The resulting normalized vector $|v_1\rangle$ can then be used as the starting vector in a second Lanczos expansion, which is then stopped after $k$ iterations. If one denotes the resulting eigenvectors and eigenvalues of the $k$-dimensional Lanczos matrix by $|f_i\rangle$ and $\epsilon_i$, then

$$3\sum_{i=1}^{k} |\langle f_i|S_z(\vec{q}_{\pi\pi})|g.s.\rangle|^2 \delta(\omega - \epsilon_i), \tag{8}$$

viewed as a distribution in $\omega$, reproduces the lowest $2k-1$ moments of the exact distribution given in Eq. (4) [10]. Thus the broad outline of the response function is determined after a few iterations, with finer details emerging as the addition of higher moments increases the resolution. The Lanczos moments technique for inclusive response functions is thus exact in two senses: the lowest $2k-1$ moments are correctly determined, and for any specified limit of resolution (e.g., that achieved in some experiment) the iterations can be continued until a sufficient number of moments are obtained to produce an overall profile that is exact at the scale.

Figs. 1a and 1b give the dynamic spin response function per spin, $S(\vec{q}_{\pi\pi}, \omega)/2L$ for $L = 12$, $J_\perp/J_\parallel = 1$, and $J_\parallel = 1$, smoothed by a Gaussian resolution function with a standard deviation of 0.05. These initial calculations were done to determine, for this choice of resolution, the required number of iterations. These results, and those of Fig. 2, suggest that $\sim 70$ iterations are needed to produce a fully converged distribution. This conclusion, however, depends on one's choice of $J_\perp/J_\parallel$: when this ratio is increased, the proportion of the response carried by high-lying excitations drops, while at the same time the distribution of strength at high $\omega$ exhibits more structure. Both effects slow the rate of convergence. Thus we found it necessary to use 200 iterations in the case of $J_\perp/J_\parallel = 5.0$.

A second numerical issue is the absence of exact orthogonality of the Lanczos vectors when Eq. (2) is implemented numerically [11]. Errors associated with the overlaps of a newly generated Lanczos vector $|v_i\rangle$ with previous vectors can be quite troublesome: spurious overlaps with extremum eigenvectors can grow, in successive iterations, because they contribute so strongly to higher moments. This can lead, for example, to repeated reconvergence of the ground state and to distortions in inclusive response functions. In many applications this difficulty makes repeated reorthogonalization by the Gramm-Schmidt procedure necessary, a step that becomes costly when a large number of iterations are performed. However, the need for reorthogonalization varies greatly from application to application. The results from our exploration of this issue for Heisenberg spin ladders is shown in Fig. 2, where a $k = 70$ calculation with reorthogonalization in each iteration is compared to $k = 70$ and 500 calculations without. The calculations were performed for $L = 10$ and $J_\perp/J_\parallel = 1$. The Heisenberg ladder Hamiltonian appears to be remarkably immune to numerical orthogonality difficulties: no differences among the three calculations are readily discernable. Thus the remainder of the calculations reported here were done without a reorthogonalization step.

Table III and Figs. 3a-l give our main results, $S(\vec{q}_{\pi,\pi},\omega)$ per spin for $J_{\parallel} = 1.0$ and $J_{\perp}/J_{\parallel}$ ranging from 0.0 to 5.0. The distributions have been smoothed by a Gaussian resolution function with $\sigma = 0.05$. As the momentum transfer corresponds to the inverse lattice size, the operator reverses the orientation of nearest neighbor spins: a low-lying spin triplet state increasingly dominates the spin response function as $J_{\perp}$ is increased (see Table III). The gap between the singlet ground state and the strong triplet state increases with increasing $J_{\perp}/J_{\parallel}$, in agreement with the strong coupling (large $J_{\perp}$) prediction of $E_{gap} \sim J_{\perp} - J_{\parallel}$.

The strength above the first triplet state is always modest, starting at $\sim 16$ % for $J_{\perp}/J_{\parallel} = 0$ and declining montonically to $\sim 0.02$ % for $J_{\perp}/J_{\parallel} = 5$. The pattern of this strength, however, becomes more distinctive with increasing $J_{\perp}/J_{\parallel}$. Thus in principle this part of the dynamic spin response, while accounting for little of the total strength, could be used in combination with the singlet-triplet gap to test whether real materials respond as simple spin ladders.

The total response strength per spin is not a monotonic function of $J_{\perp}/J_{\parallel}$, but instead increases from $J_{\perp}/J_{\parallel} = 0$ to a peak at about $J_{\perp}/J_{\parallel} \sim 0.5$, then declines steadily above this value. The results are shown in Table III. We also examined the evolution of the total strength, and strength carried by the first triplet state, as a function of $L$ for fixed $J_{\perp}/J_{\parallel} = 1.0$. The fraction of strength carried by the first triplet state appears to converge rapidly: the results for $L = 6$, 8, 10, and 12 are 97.66%, 97.04%, 96.81%, and 96.73%, which extrapolate to a bulk limit of 96.70%. The bulk limit of the absolute strength per spin carried by the first triplet state is somewhat less certain, as this quantity appears to be rather slowly approaching an asymptote in our calculations: the results for $L = 6$, 8, 10, and 12 are 3.26745, 3.52202, 3.66170, and 3.73619. These values can be fit very well by a scaling function of the form of Eq. (3) with $p = 0$, leading to a bulk limit of 3.83. To compare with [8], we repeated these calculations for open boundary conditions, finding 2.80696, 2.99800, 3.12141, and 3.20260. The $L = 8$ result agrees with the corresponding exact calculation of [8], who also used their approximate plaquette basis calculations to estimate a bulk limit open-boundary-condition value of 3.30-3.36. (Note that the values given in [8] have been multiplied by 3 to take into account the different normalization of $S(\vec{q}_{\pi\pi},\omega)$ used there.) If we employ the same scaling function in this case, our calculations yield a bulk limit 3.35, compatible with this range. Unfortunately we have results for too few $L$ values to test numerically whether our assumed scaling function is reasonable. Thus there could be substantial errors in these estimates.

Finally, we stress that the dynamical spin response calculations described here are quite practical on standard workstations. Each of the graphs comprising Fig. 3 required about two hours of CPU time on a large-memory (1Gb) DEC Alpha 500. The stability of Heisenberg spin ladder calculations in the absence of reorthogonalization contributes to the numerical efficiency: reorthogonalization can be costly because of the need for Lanczos vector storage and associated i/o operations. Thus such methods-of-moments calculations of dynamical spin responses for Heisenberg spin chains should be practical for $L = 14$ on workstations similar to ours, and extensions well beyond this possible with supercomputers.

We thank J. Piekarewicz and J. Shepard for bringing this problem to our attention and for several helpful discussions. This work was supported in part by the US Department of Energy. One of us (DY) ackowledges the support of the University of Washington and National Science Foundation Physics Research Experiences for Undergraduates Program.

FIGURES

FIG. 1. Results for $2 \times L = 24$ spin sites showing the convergence of the dynamical spin response per spin as a function of the number of iterations performed. The required number of iterations depends on the desired resolution, which in these calculations is determined by the choice of smearing function. A Gaussian with $\sigma = 0.05$ has been used.

FIG. 2. The dynamical spin response per spin for a ladder with $2 \times L = 20$ sites calculated with (70 iterations) and without (70 and 500 iterations) reorthogonalization. The results without reorthogonalization remain stable well past the point where $S(\vec{q}_{\pi,\pi}, \omega)$ has fully converged.

FIG. 3. The evolution of the spin response function per site for ladders with $2 \times L = 24$ sites as a function of $J_\perp / J_\parallel$.

TABLES

TABLE I. The ground state energy per spin and interaction strength $E_0/2LJ_{||}$ for $2 \times L$ Heisenberg spin ladders with $J_\perp = J_{||}$.

| L | Present | Barnes *et al.* |
|---|---|---|
| 4 | -0.6025112 | -0.602511 |
| 5 | -0.5638793 | |
| 6 | -0.5844372 | -0.584437 |
| 7 | -0.5739430 | |
| 8 | -0.5802030 | -0.580203 |
| 9 | -0.5766331 | |
| 10 | -0.5788595 | -0.578860 |
| 11 | -0.5775071 | |
| 12 | -0.5783722 | -0.578375 |
| 13 | -0.5778259 | |
| 14 | -0.5781816 | |

TABLE II. As in Figure 1, only for the singlet-triplet gap $(E_1 - E_0)/J_{\|}$.

| $L$ | Present | Barnes *et al.* |
|---|---|---|
| 4 | 0.8200894 | 0.820089 |
| 5 | 0.8761249 | |
| 6 | 0.6265690 | 0.626570 |
| 7 | 0.7734289 | |
| 8 | 0.5573976 | 0.557398 |
| 9 | 0.7039126 | |
| 10 | 0.5281070 | 0.528106 |
| 11 | 0.6558908 | |
| 12 | 0.5147836 | 0.514999 |
| 13 | 0.6218955 | |
| 14 | 0.5084957 | |

TABLE III. The dynamical spin response per spin for the $2 \times L = 24$ Heisenberg spin ladder divided into the lowest triplet state contribution and the contribution carried by all higher states.

| $J_\perp/J_\parallel$ | lowest $S = 1$ state | higher states |
|---|---|---|
| 0.0 | 2.484 (84.3%) | 0.464 |
| 0.2 | 3.635 (89.1%) | 0.443 |
| 0.4 | 4.148 (91.7%) | 0.376 |
| 0.6 | 4.219 (93.8%) | 0.281 |
| 0.8 | 4.033 (95.4%) | 0.193 |
| 1.0 | 3.736 (96.7%) | 0.126 |
| 1.2 | 3.431 (97.7%) | 0.081 |
| 1.4 | 3.162 (98.4%) | 0.053 |
| 1.6 | 2.940 (98.8%) | 0.035 |
| 1.8 | 2.758 (99.2%) | 0.023 |
| 2.0 | 2.610 (99.4%) | 0.016 |
| 5.0 | 1.864 (99.98%) | 0.0004 |

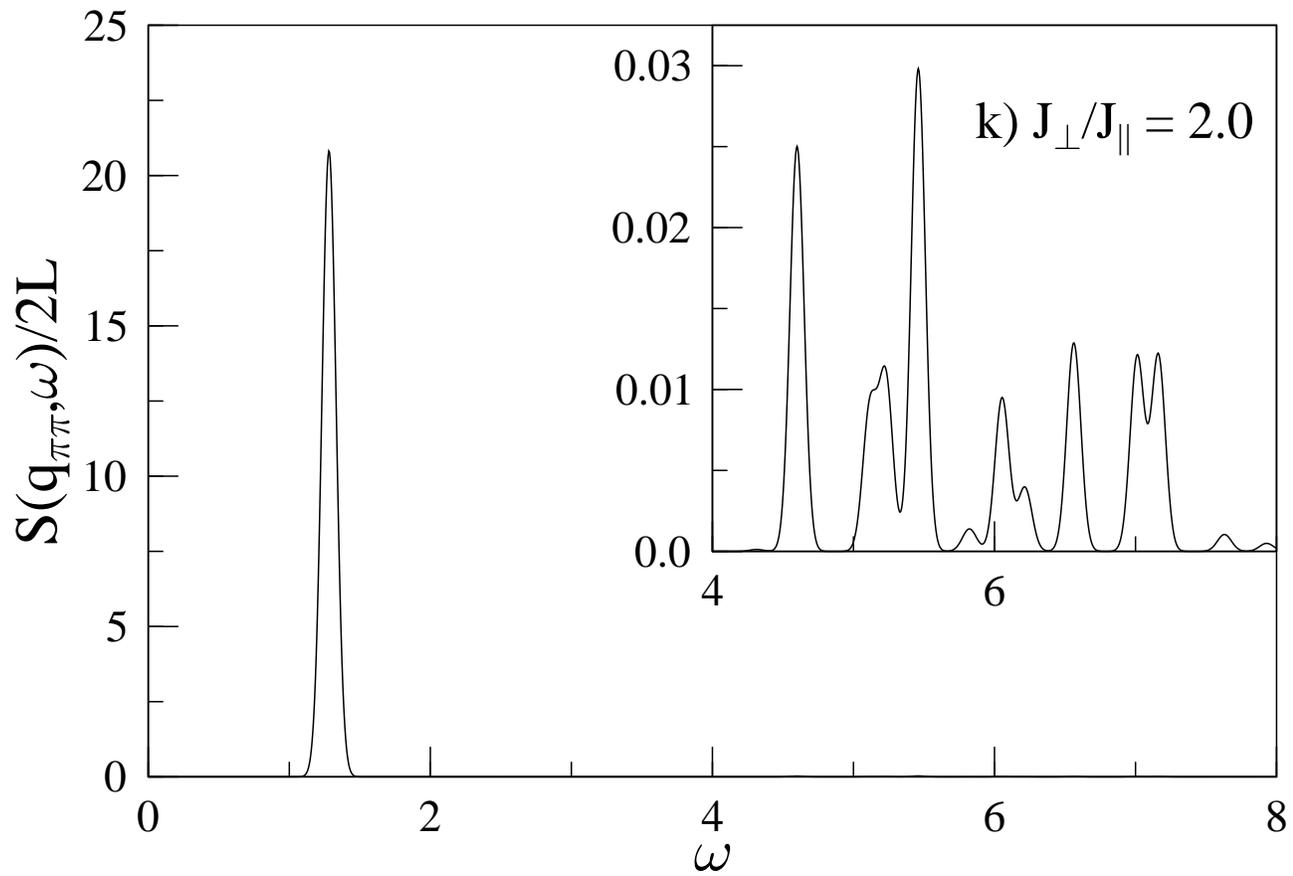
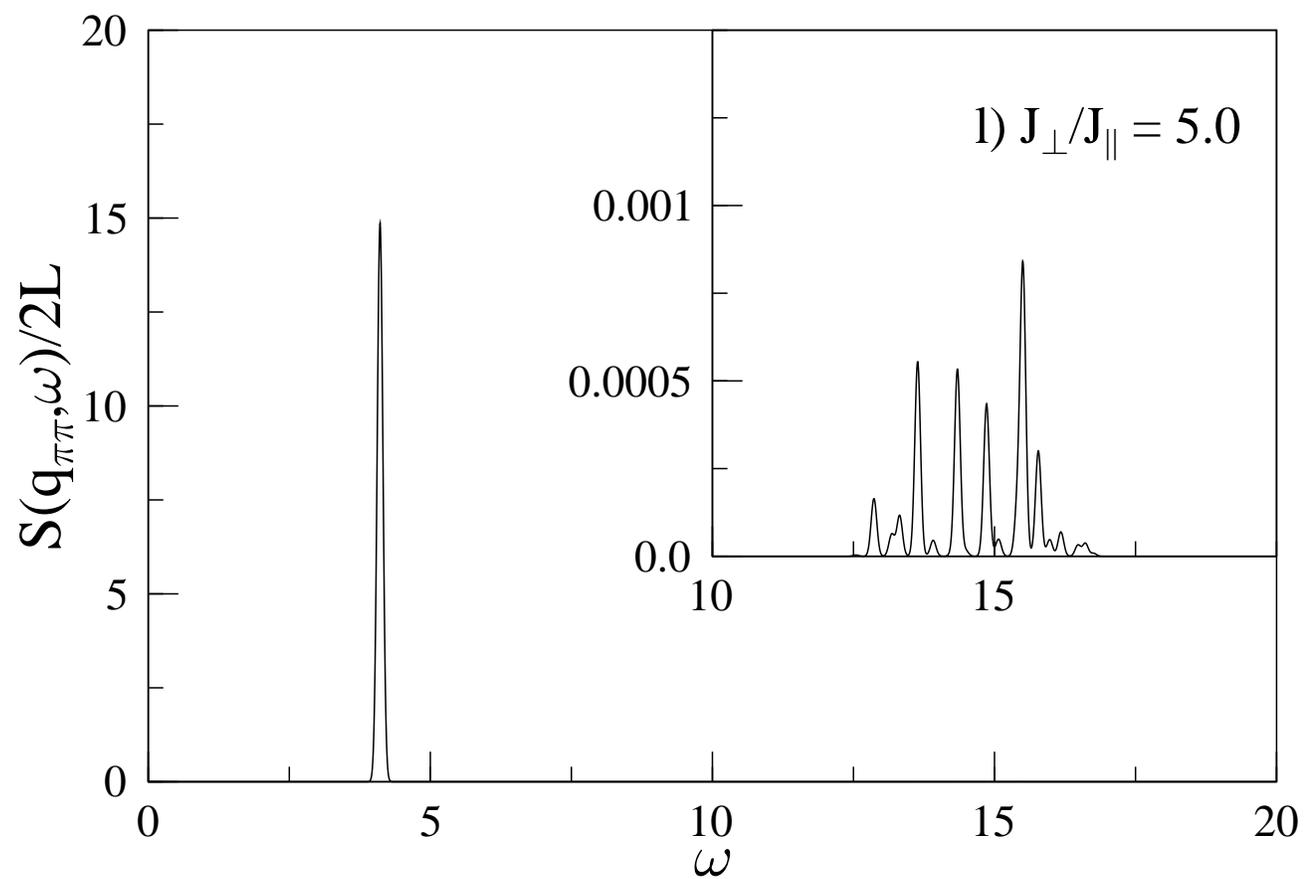

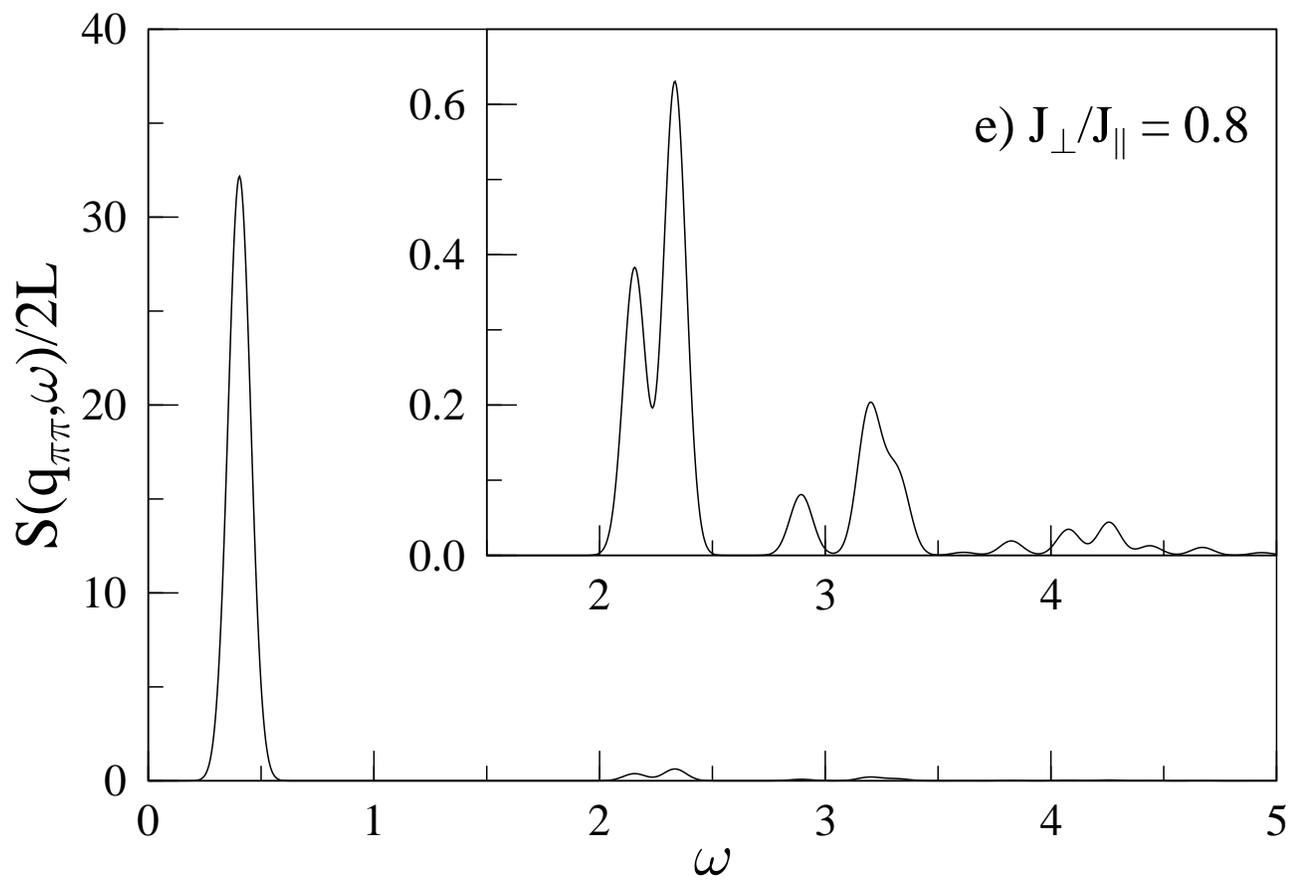

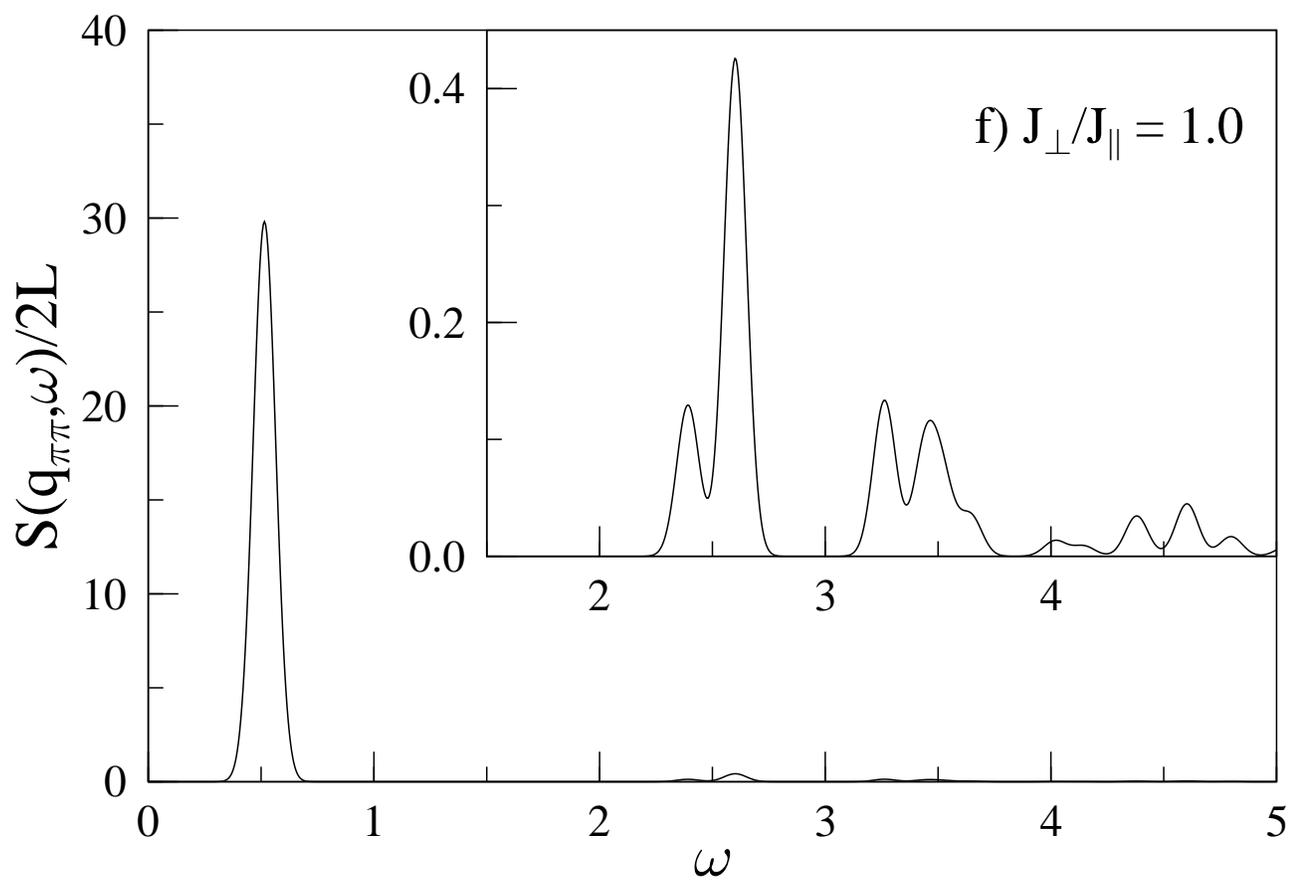

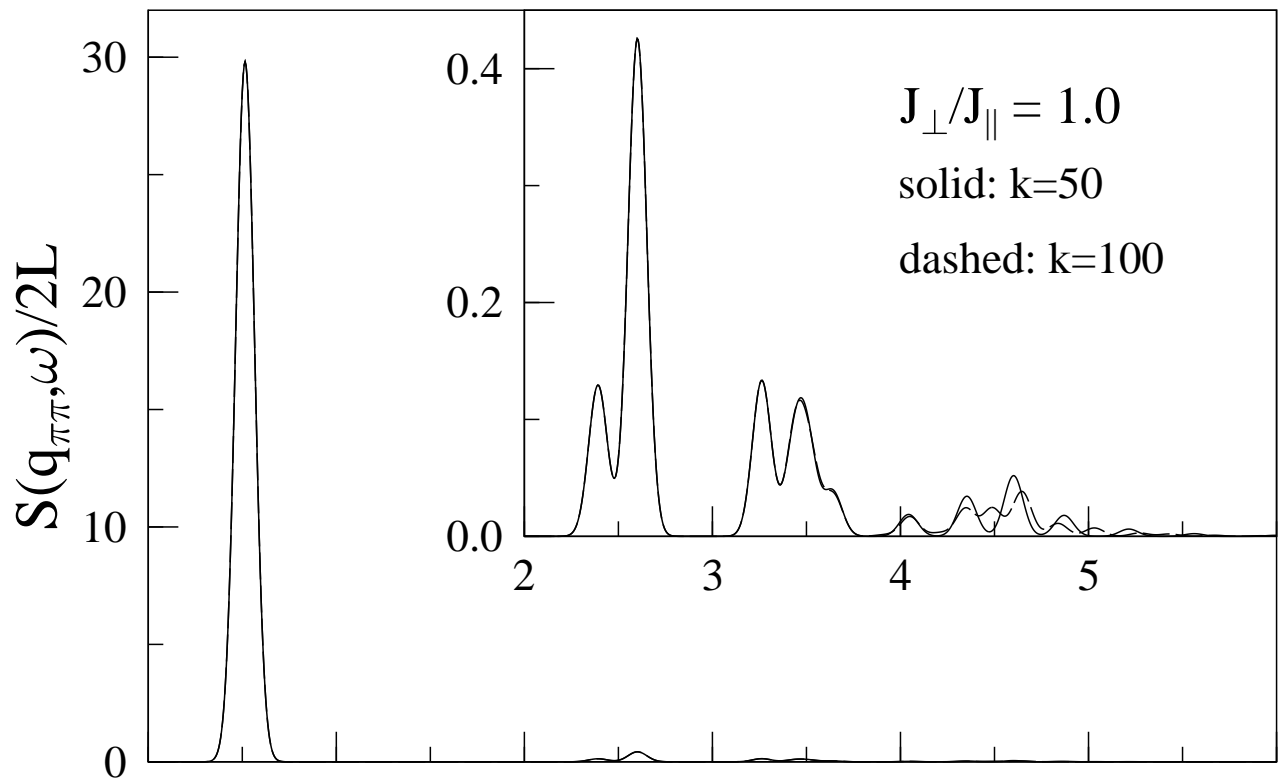
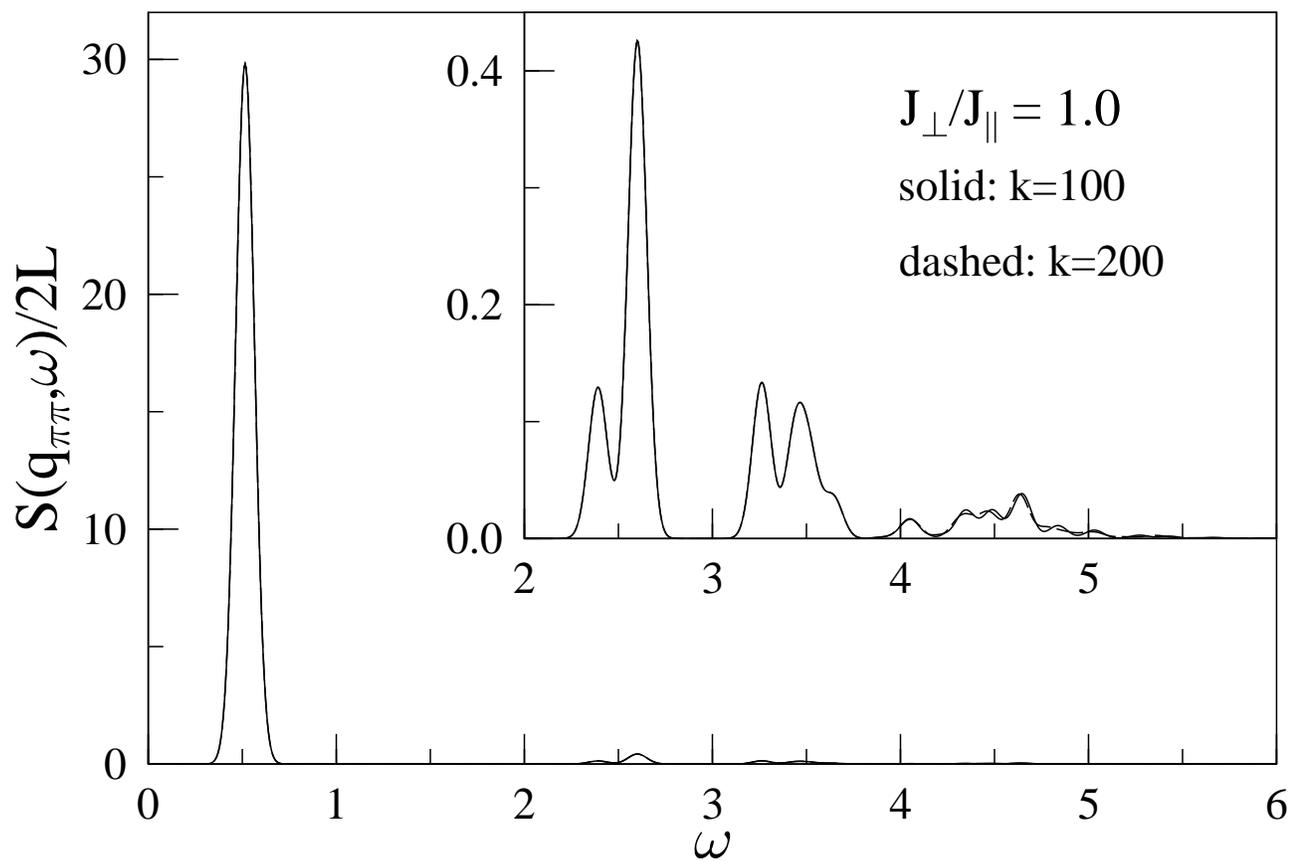

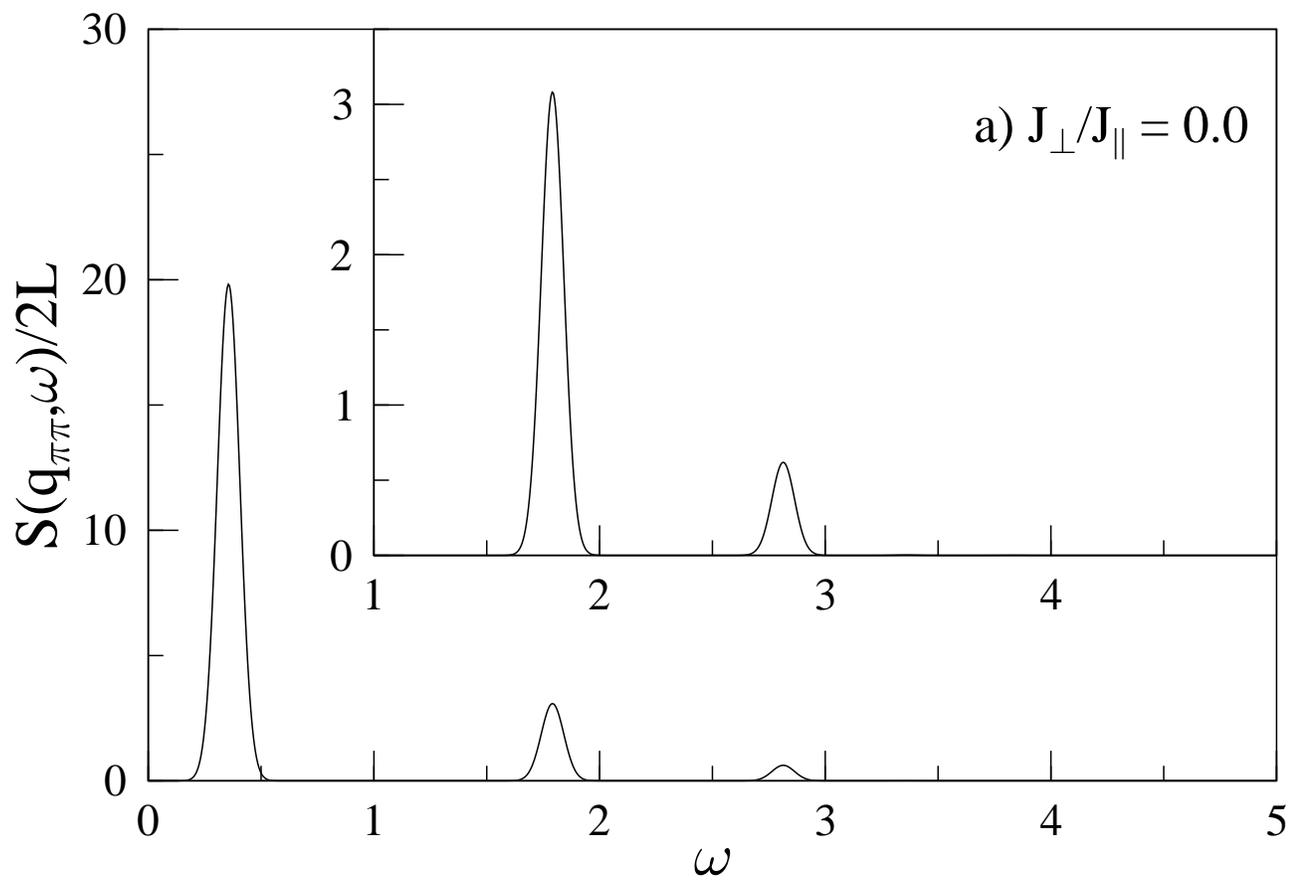
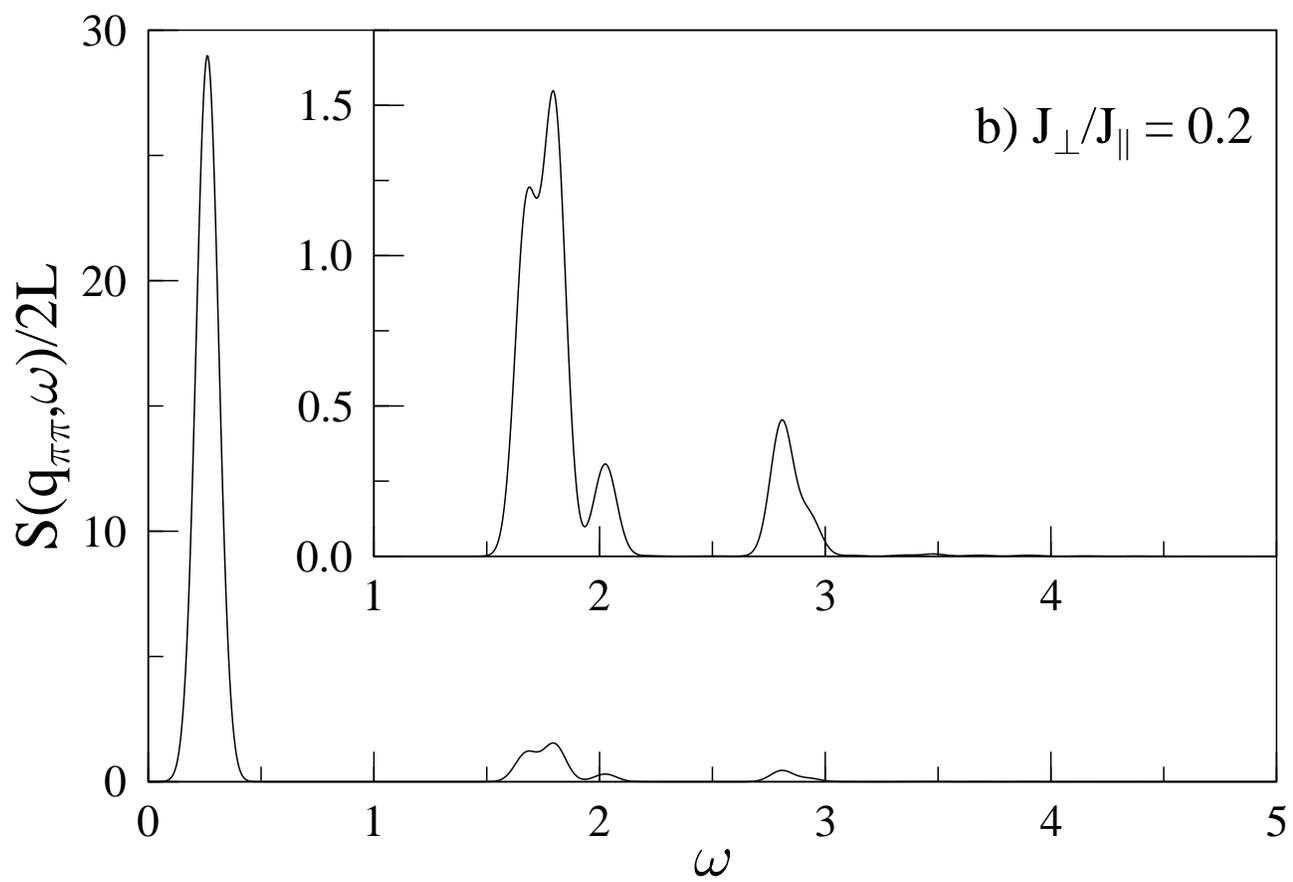

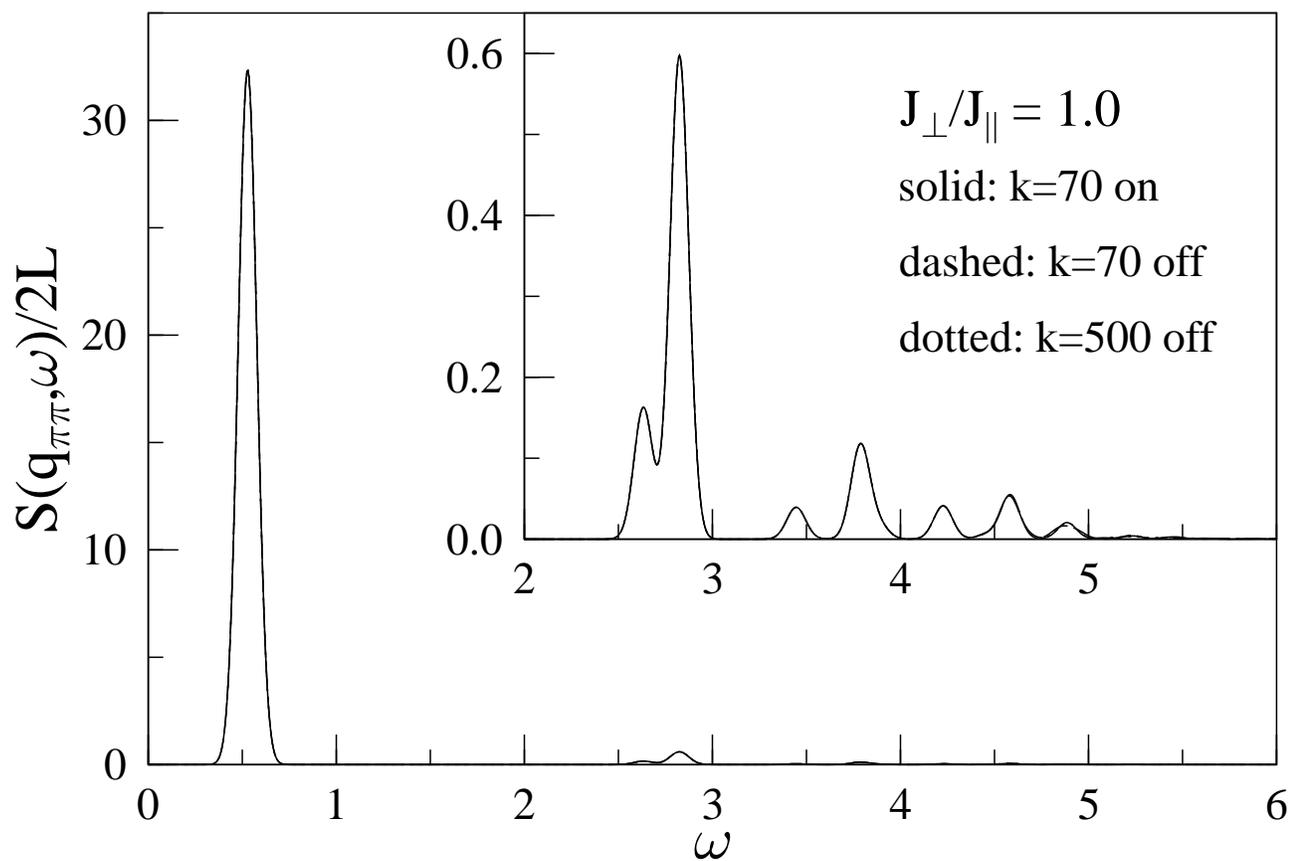

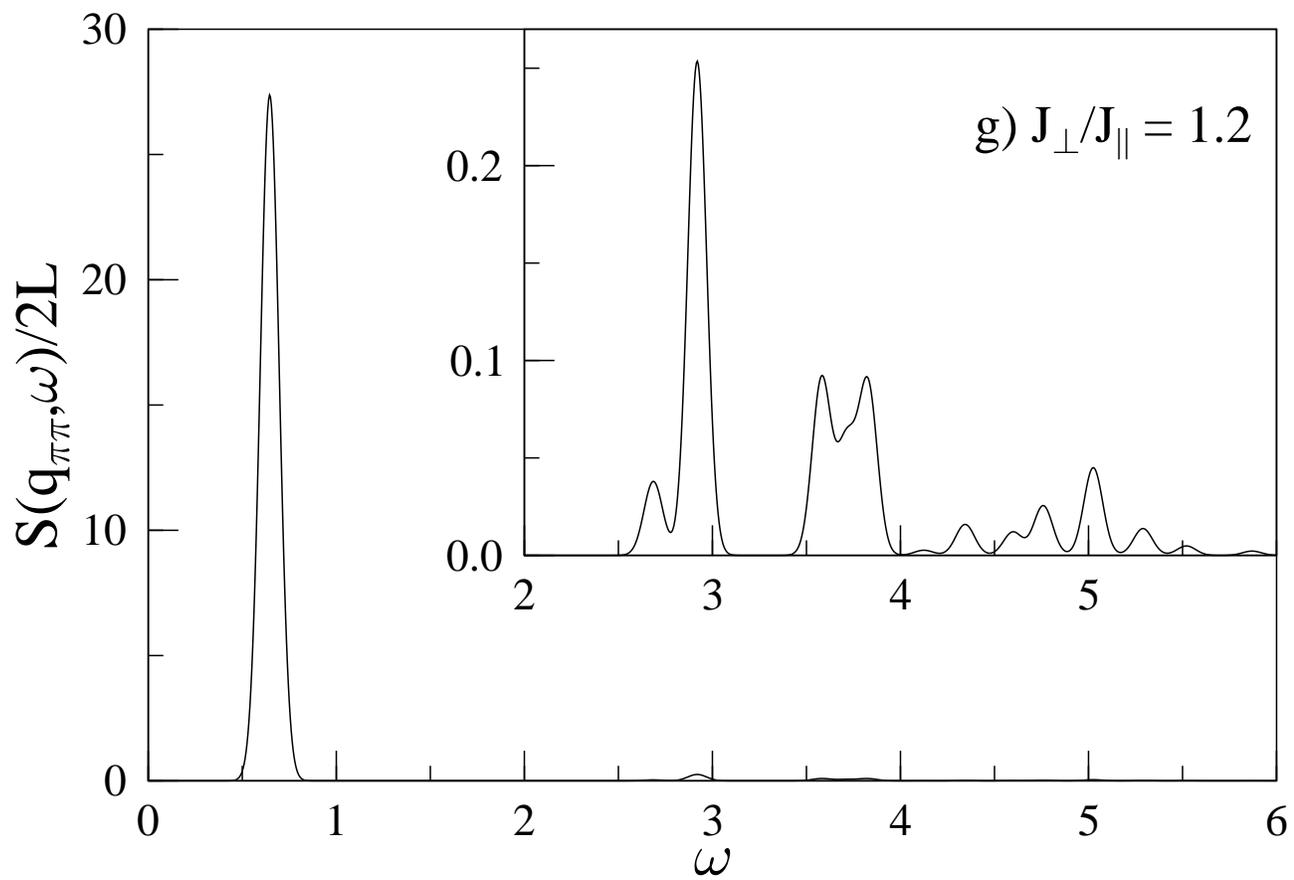
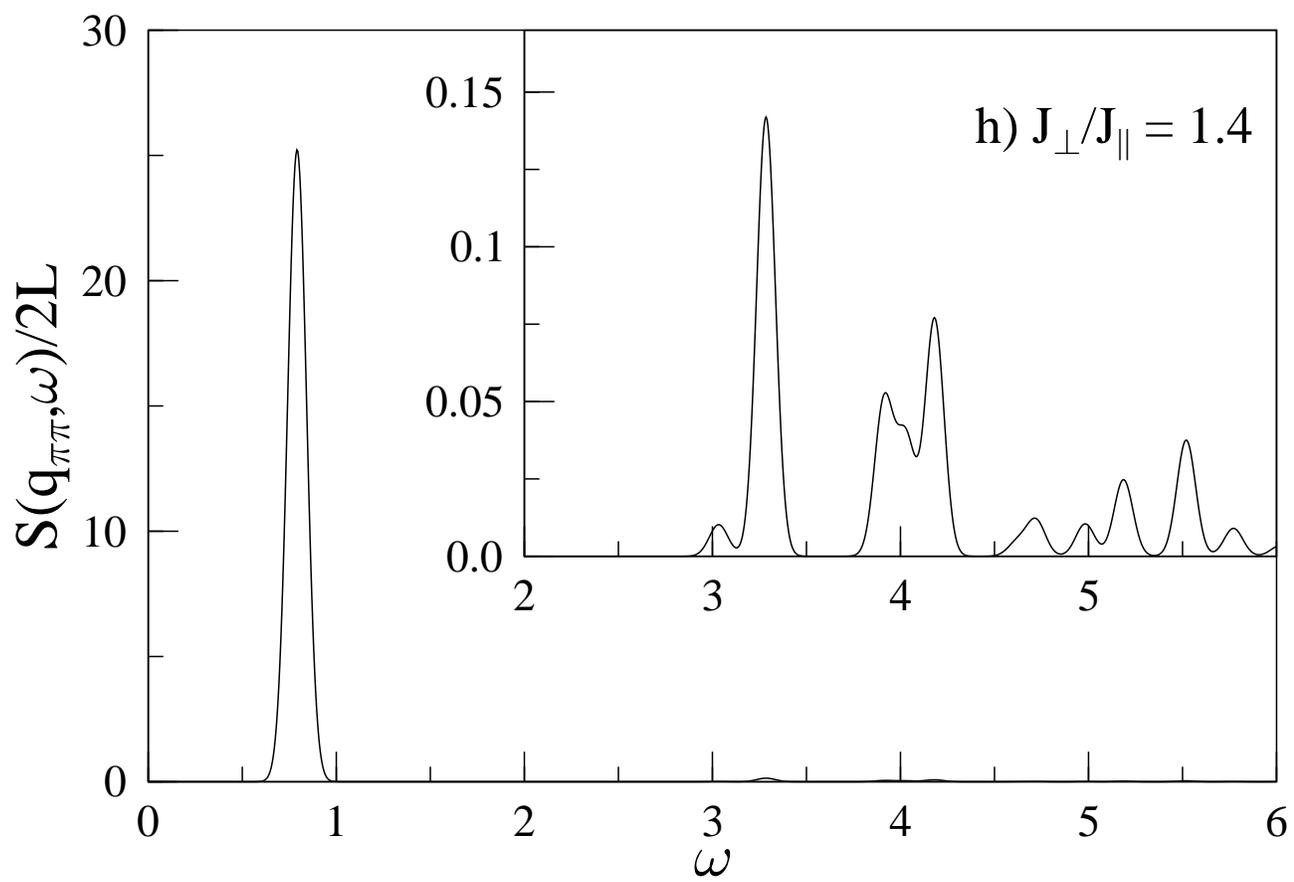

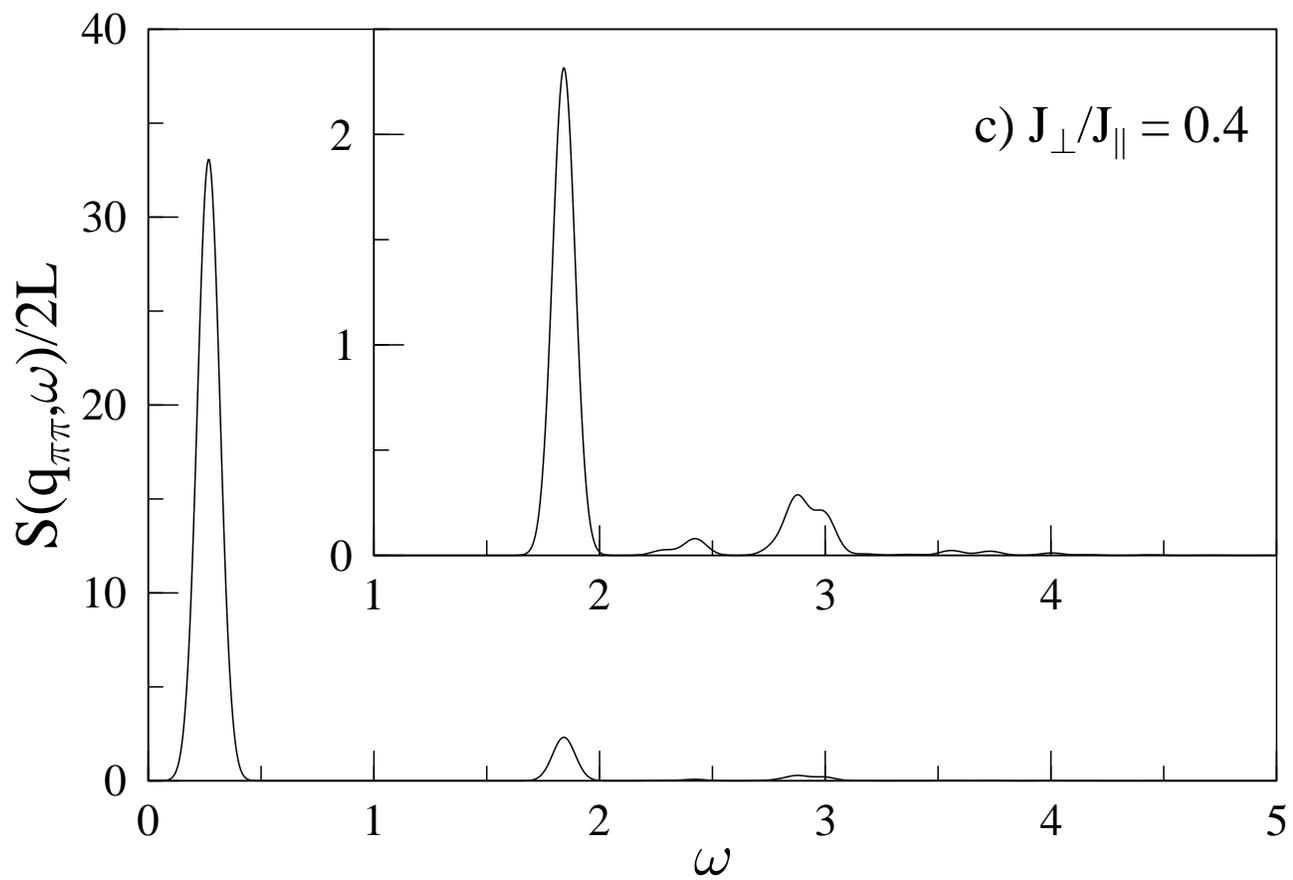

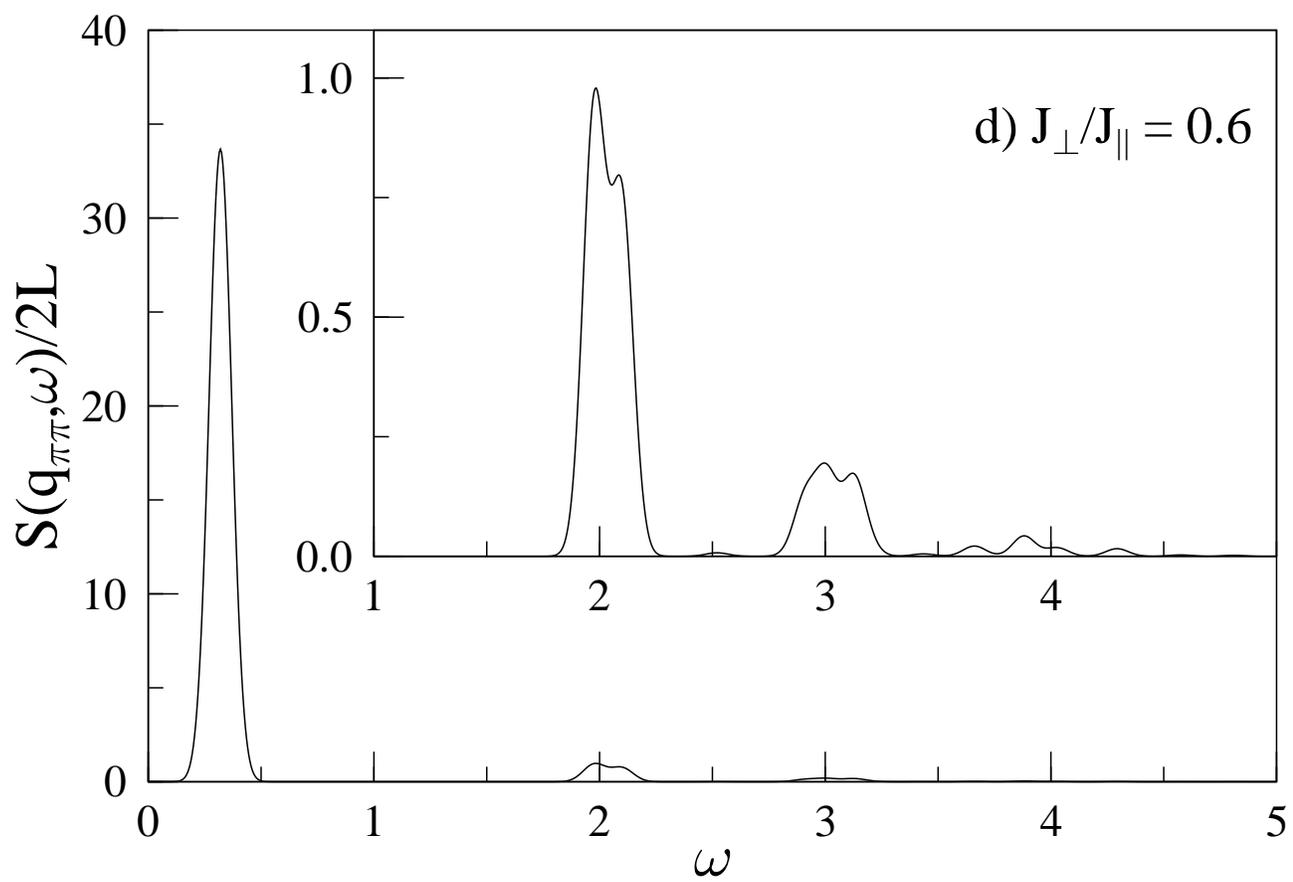

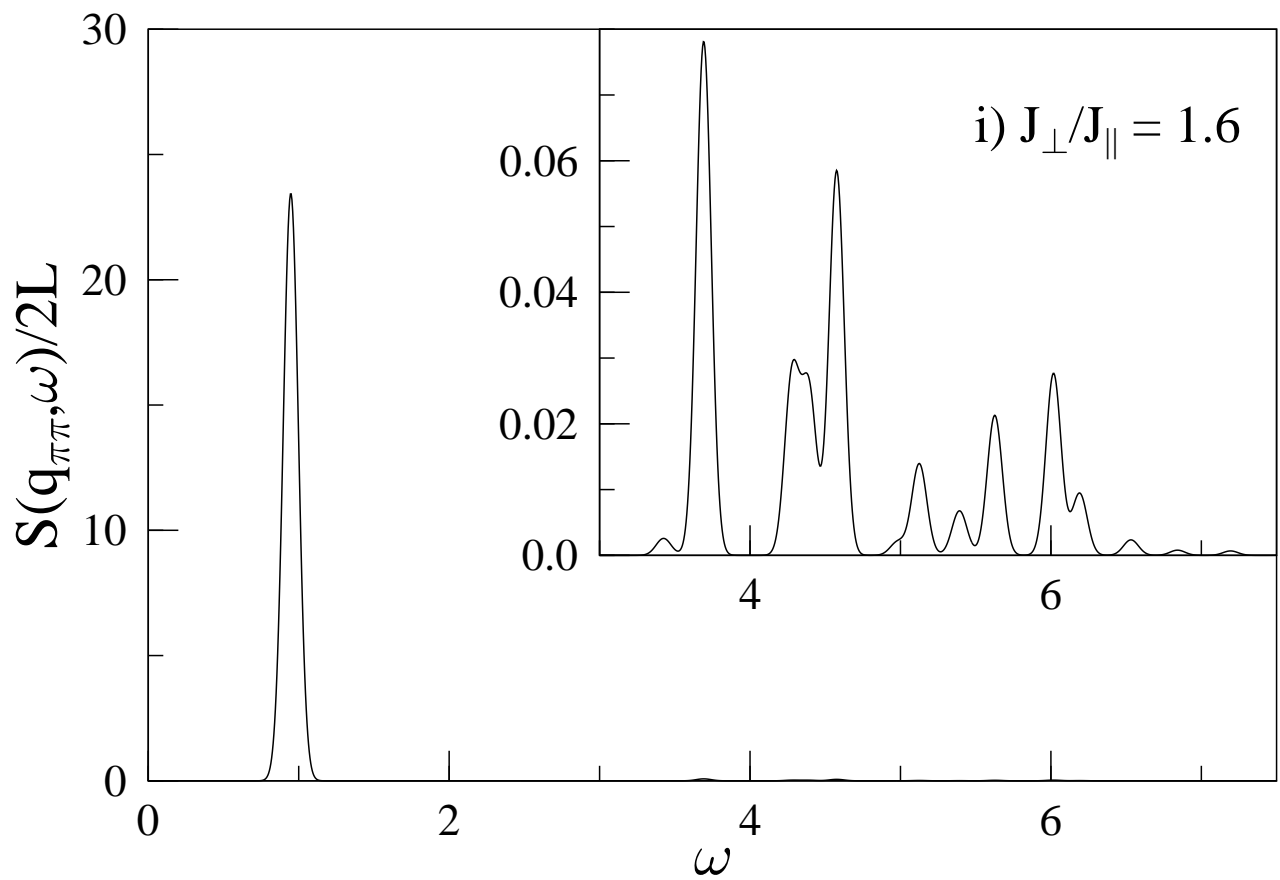
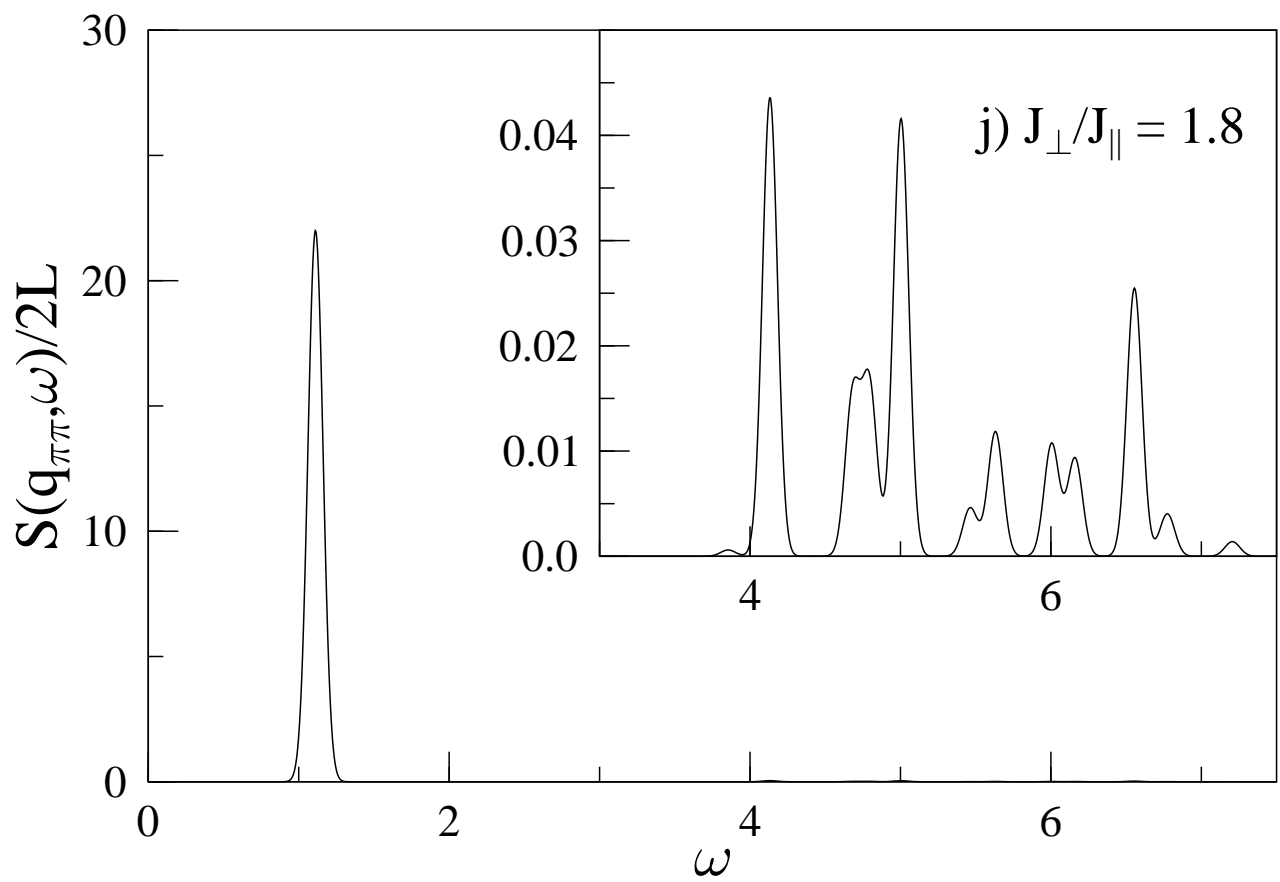